\documentclass[english,aps,floats,onecolumn,showpacs,nofootinbib]{revtex4}
\usepackage{pslatex}
\usepackage[T1]{fontenc}
\usepackage[latin1]{inputenc}
\usepackage{graphicx}
\usepackage{epsfig}

\usepackage{calc}
\usepackage{ifthen}

{
{
{
\newcommand{\bea}{\begin{eqnarray}}
\newcommand{\eea}{\end{eqnarray}}

\newcommand{\nc}{\newcommand}
\nc{\renc}{\renewcommand}
\nc{\eqs}[2]{\mbox{Eqs.~(\ref{#1},\,\ref{#2})}}
\nc{\eq}[1]{\mbox{Eq.~(\ref{#1})}}
\nc{\figs}[2]{\mbox{Figs.~(\ref{#1},\,\ref{#2})}}
\nc{\fig}[1]{\mbox{Fig~.(\ref{#1})}}
\nc{\be}[1]{\begin{equation} \mbox{$\label{#1}$}}
\nc{\ee}{\vspace{0.1cm}\end{equation}}

\newcommand{\bean}{\begin{eqnarray*}}
\newcommand{\eean}{\end{eqnarray*}}

\def\bfx{{\bf x}}
\def\bfk{{\bf k}}
\def\bfl{{\bf l}}

 \def\gae{\; ^{>}_{\sim} \;}

\begin{document}
\title{Isocurvature and Curvaton Perturbations with Red Power Spectrum and Large Hemispherical Asymmetry}
\author{John McDonald}
\email{j.mcdonald@lancaster.ac.uk}
\affiliation{Lancaster-Manchester-Sheffield Consortium for Fundamental Physics, Cosmology and Astroparticle Physics Group, Dept. of Physics, University of 
Lancaster, Lancaster LA1 4YB, UK}
\begin{abstract}

      We calculate the power spectrum and hemispherical asymmetry of isocurvature and curvaton perturbations due to a complex field $\Phi$ which is evolving along the tachyonic part of its potential. Using a semi-classical evolution of initially sub-horizon quantum fluctuations, we compute the power spectrum, mean field and hemispherical asymmetry as a function of the number of e-foldings of tachyonic growth $\Delta N$ and the tachyonic mass term $cH^2$. We find that a large hemispherical asymmetry due to the modulation of $|\Phi|$ can easily be generated via the spatial modulation of $|\Phi|$ across the horizon, with $\Delta |\Phi|/|\Phi| > 0.5$ when the observed Universe exits the horizon within 10-40 e-foldings of the beginning of tachyonic evolution and $c$ is in the range 0.1-1. The spectral index of the isocurvature and curvaton perturbations is generally negative, corresponding to a red power spectrum. Dark matter isocurvature perturbations due to an axion-like curvaton with a large hemispherical asymmetry may be able to explain the hemispherical asymmetry observed by WMAP and Planck. In this case, the red spectrum can additionally suppress the hemispherical asymmetry at small scales, which should make it easier to satisfy scale-dependence requirements on the asymmetry from quasar number counts.

\end{abstract}
\maketitle

\section{Introduction}

       Isocurvature perturbations are a natural feature of a range of dark matter and baryogenesis models. A common class of model is based on a complex field $\Phi$ whose phase determines the density of the isocurvature component, with axion dark matter and Affleck-Dine baryogenesis being two well-known examples.  In these models the isocurvature perturbations are associated with quantum fluctuations of the phase angle. 

       The nature of the isocurvature perturbations depends on the evolution of $\Phi$ along its potential and the time at which the observed Universe exits the horizon. In this paper we will explore the case where $\Phi$ evolves from $|\Phi| = 0$ along the tachyonic part of its potential, where the potential is of the form $V(\Phi) \approx -cH^{2} |\Phi|^2$. In particular, we will compute the 
mean field in a horizon volume, the spectrum of isocurvature perturbations and the mean hemispherical asymmetry across a horizon volume, as a function of the time of horizon exit and of $c$. We will show that a red isocurvature perturbation spectrum and a significant hemispherical asymmetry are generic features of this class of isocurvature perturbation model. 

    A topical application of these results is to the hemispherical power asymmetry of the CMB, observed initially by WMAP \cite{wmap} and recently confirmed by Planck \cite{planck,pesky}. One possible explanation is a dark matter isocurvature perturbation with a large hemispherical asymmetry due to the decay of a curvaton \cite{kam}. (See also \cite{kam2,kam3,nb,lyth})\footnote{Recent alternative proposals to account for the asymmetry are discussed in \cite{lingfei,liu}.}.  We can apply our isocurvature analysis directly to the case of an axion-like curvaton due to a complex field $\Phi$. A modulation across the horizon by $\Delta |\Phi|/|\Phi| \gae 0.5$  is then necessary to account for the observed asymmetry \cite{kam}. We will determine the conditions under which such a large modulation can be achieved. The red curvaton perturbation spectrum expected in these models will alter the analysis of the model from previous studies, which assume a scale-invariant curvaton spectrum \cite{kam}. In particular, a red spectrum should allow the model to more easily satisfy scale-dependence requirements on the hemispherical asymmetry from quasar number counts \cite{quasar} and other observations \cite{don}.

     In Section 2 we introduce the model and the method we use for its analysis. In Section 3 we derive the mean $\Phi$ field in a horizon volume and the power spectrum and hemispherical asymmetry of the isocurvature perturbations. In Section 4 we apply our analysis to obtain the hemispherical asymmetry and power spectrum in specific cases. In Section 5 we present our conclusions. Some details of the semi-classical method are included in an Appendix. 

     \section{Evolution of Isocurvature Fluctuations in a Tachyonic Potential} 

  In our model we consider a complex field $\Phi = \frac{\phi}{\sqrt{2}}e^{i \theta}$ to be the origin of the isocurvature perturbations. (We will refer to $\Phi$ as the 'isocurvature field' in the following.) The density of the isocurvature component of the Universe is assumed to be proportional to $\theta^{n}$ for some power $n$. For example,  $n=2$ in the case of axion-like dark matter in the small misalignment angle limit, while $n = 1$ for baryogenesis in the case where the CP-violating phase is proportional to $\theta$, as in the case of Affleck-Dine baryogenesis. 

    A closely-related example, which is relevant to the WMAP/Planck hemispherical asymmetry, is a dark matter isocurvature perturbation due to the decay of a sub-dominant curvaton field \cite{kam}. In the case of an axion-like curvaton field, the curvaton is proportional to $\theta$. 

     The potential of the isocurvature field is 
\be{e1}   V(\Phi) = -cH^{2} |\Phi|^{2}   +  V_{lift}(|\Phi|)   ~.\ee
Here $c$ parameterizes the tachyonic mass term. $V_{lift}$ contains the terms which determine the minimum of the potential $|\Phi|_{min}$. These terms will not play a role in our analysis.

   We assume that the isocurvature field is initially at $|\Phi| = 0$, for example by $c$ being initially negative. The field then begins to evolve from $|\Phi| = 0 $ at an initial time which we define to be $t = 0$, after which $c$ is positive and the field evolves in the resulting tachyonic potential. Although we do not single out any particular class of model, we note that this scenario is a natural possibility in the context of supergravity. A $-cH^2 |\Phi|^2$ term is a natural supergravity correction and $c$ can change sign if the dominant inflaton field changes at some time, with the later inflaton field having different K\"ahler couplings to the isocurvature field.

   The field $\Phi$ can be expanded in terms of real fields,   $\Phi = (\phi_{1} + i \phi_{2})/\sqrt{2}$. Since the evolution 
is linear during the tachyonic regime, it is the same for $\phi_{1}$ and $\phi_{2}$, so we will consider a generic real scalar field $\phi$.
      The initial state of the field at $t = 0$ is a Bunch-Davies vacuum on subhorizon scales. We will assume that no superhorizon perturbations can form before the tachyonic growth era, which is true if $|c| > 1$ before the onset of tachyonic growth.

In order to evolve the field from this initial state, we will use a semi-classical method that is commonly used in numerical simulations of tachyonic preheating \cite{felder,rajantie,mbs,latticeeasy}. The method is based on the Wigner function for the field. 
The Wigner function method replaces the quantum field with a 
classical field which evolves from initial conditions which have a classical probability distribution. Quantum correlation functions in the semi-classical limit are then equal to the classical correlation functions averaged over the classical initial probability distribution. We review the method in the Appendix. 

     The initial classical modes are Gaussian distributed with a random phase. Although the initial conditions are sub-horizon and not initially semi-classical, the modes can still be evolved from classical initial conditions, provided that the mode enters the semi-classical regime while the field is still undergoing linear evolution i.e. the field is still dominated by the tachyonic mass term \cite{latticeeasy,mbs}. This is because the Wigner function in the semi-classical limit becomes a classical phase space density which must reproduce the late-time semi-classical behaviour. Therefore the final result for late-time semi-classical modes will be the same, regardless of when the classical initial conditions are applied.   

The rms values of the initial classical modes in de Sitter space are (Appendix) 
\be{e14}   |q(\bfk, \eta)|_{rms} = \frac{1}{\sqrt{2 k}} 
\left(1 + \frac{H^{2}}{k^{2}}\right)^{1/2}   ~\ee 
and
\be{e15}   |p(\bfk, \eta)|_{rms} = \frac{H}{\sqrt{2 k}} 
\left(1 + \frac{H^{2}}{k^{2}}\right)^{-1/2}   ~,\ee
where $\eta = -1/aH$ is the conformal time and initially $\eta = -1/H$.  

  Since we will be interested in mean quantities, we will evolve the classical modes from the RMS initial conditions \eq{e14} and \eq{e15}, with the initial time defined to be $t = 0$ and the initial scale-factor defined to be $a = 1$. The random phases will not play a role in our analysis.

   The classical modes during the tachyonic era are defined by 
\be{e16}  \phi(\bfx, a) = \frac{1}{\sqrt{V}} \sum_{\bfk}  \phi_{\bfk}(a) e^{i \bfk.\bfx}    ~,\ee  
where we use the scale factor to parameterize time. (We define modes in a box of volume $V$ and side $L$, taking the continuum limit where appropriate.)  
The mode equation is 
\be{e17} \frac{\partial^{2} \phi_{\bfk}}{\partial a^2} + \frac{4}{a} \frac{\partial \phi_{\bfk}}{\partial a} + \frac{k^{2}}{a^{4} H^{2}} \phi_{\bfk} = \frac{c}{a^{2}} \phi_{\bfk}     ~.\ee 
The general solution is
\be{e18}  \phi_{\bfk}(a) = \frac{c_{1}}{a^{3/2}} J_{-\frac{1}{2}\sqrt{4 c + 9}}\left(\frac{k}{aH}\right) + 
\frac{c_{2}}{a^{3/2}} J_{\frac{1}{2}\sqrt{4 c + 9}}\left(\frac{k}{aH}\right)    ~,\ee 
where $J_{\nu}(x)$ are Bessel functions. 
To fix $c_{1}$ and $c_{2}$ we match this to the rms initial conditions, \eq{e14} and \eq{e15}, at $a = 1$. 
Modes corresponding to  the observed Universe will be subhorizon at $a = 1$. In this case $k > 2 \pi H$ and 
we can use the asymptotic form of the Bessel function, 
\be{e19} J_{\nu}(x) = \sqrt{\frac{2}{\pi x}} \cos\left(x - \frac{\nu \pi}{2} - \frac{\pi}{4} \right) + O\left( \frac{1}{x^{3/2}} \right) \;\;\;\;;\;\;\;\; x \gg 1   ~.\ee  
Therefore,  for $k/aH \gg 1$, which is generally true for modes at $a=1$, 
\be{e22} \phi_{\bfk}(a) = c_{1} \sqrt{\frac{2 a H}{\pi k}} \cos \left( \frac{k}{aH} + \frac{\sqrt{4 c + 9} \pi}{4} - \frac{\pi}{4} \right) + 
c_{2} \sqrt{\frac{2 a H}{\pi k}} \cos \left( \frac{k}{aH} - \frac{\sqrt{4 c + 9} \pi}{4} - \frac{\pi}{4} \right) ~.\ee 
$\phi(\bfx, a) = y(\bfx, -1/H)$ and $\dot{\phi}(\bfx, a) = \pi(\bfx, -1/H)$ at $a = 1$ and $\eta = -1/H$. Thus
$\phi_{\bfk}(1) = |q(\bfk, -1/H)|_{rms}$ and $\dot{\phi}_{\bfk}(1)  = |p(\bfk, -1/H)|_{rms}$. Therefore the initial conditions at $a = 1$ are 
\be{e20} \phi_{\bfk}(1) = \frac{1}{\sqrt{2k}} \left(1 + \frac{H^{2}}{k^{2}} \right)^{1/2}   ~\ee 
and
\be{e21} \frac{\partial \phi_{\bfk}}{\partial a}(1) = 
\frac{1}{\sqrt{2 k}} \left(1 + \frac{H^{2}}{k^{2}} \right)^{-1/2}    ~.\ee
These determines $c_{1}$ and $c_{2}$ to be 
\be{e23} c_{1} = \sqrt{\frac{\pi}{4 H}} \frac{ \left(s_{-} - 
c_{-} \frac{2 H}{k} \right)  }{ \left(c_{+}s_{-} - s_{+}c_{-} \right) }     ~\ee 
and
\be{e24}  c_{2} = - \sqrt{\frac{\pi}{4 H}} \frac{ \left(s_{+} - c_{+} \frac{2 H}{k} \right)  }{ \left(c_{+}s_{-} - s_{+}c_{-} \right) }     ~\ee 
where 
\be{e25} c_{\pm} =  \cos \left( \frac{k}{H} \pm \frac{\sqrt{4 c + 9} \pi}{4} + \frac{\pi}{4}   \right)  \;\;\;;\;\;\;  s_{\pm} =  \sin \left( \frac{k}{H} \pm \frac{\sqrt{4 c + 9} \pi}{4} + \frac{\pi}{4}   \right)    ~\ee 
and $c_{+}s_{-}-s_{+}c_{-} = -\sin(\pi \sqrt{4c+9}/2)$.

\section{Isocurvature Perturbation Spectrum and Hemispherical Asymmetry}

\subsection{Mean field}

As the modes grow, there will be a mean value of the field in a horizon-sized volume at a given $a$, due to modes which have wavelengths larger than the horizon. 
The rms field is given by 
\be{e28} <\phi(\bfx, t)^{2}> = \frac{1}{(2\pi)^{3}} \int  |\phi_{\bfk}|^{2}  d^{3}k   ~.\ee 
We define $J(k/aH)$ by
\be{n1} J\left(\frac{k}{aH}\right) = \tilde{c}_{1} J_{-\frac{1}{2}\sqrt{4 c + 9}}\left(\frac{k}{aH}\right) + 
\tilde{c}_{2} J_{\frac{1}{2}\sqrt{4 c + 9}}\left(\frac{k}{aH}\right)    ~,\ee 
where 
\be{n2}   \tilde{c}_{1,2} = \sqrt{\frac{4 H}{\pi}} c_{1,2}    ~.\ee 
Then 
\be{n2a}  \phi_{\bfk}(a) = \sqrt{\frac{\pi}{4 H a^{3}}} J\left(\frac{k}{aH}\right)   ~\ee
and
\be{n3} <\phi(\bfx, t)^{2}> = \frac{1}{8 \pi H a^{3}} 
\int_{k_{min}}^{k_{max}} J^{2}\left(\frac{k}{aH}\right)k^{2} dk    ~.\ee 
Here  $k_{min}  = 2\pi H$ corresponds to initial the horizon-sized mode at $a=1$ and $k_{max} = 2\pi a H$ is the horizon-sized mode at $a$.  Changing variable to $y = k/aH$, we obtain
\be{n4} \overline{\phi}^{2}(a)  \equiv <\phi(\bfx, t)^{2}>  = 
\frac{H^{2}}{8 \pi} K_{1}(a)   ~,\ee
where 
\be{n5}   K_{1}(a)  = \int_{\frac{2 \pi}{a}}^{2 \pi} J^{2}(y) y^{2} dy     ~.\ee  

   It is useful to have an approximation for $\overline{\phi}^{2}(a)$ which shows the $a$ dependence explicitly. To do this we split the integral into two parts, 
\be{n6} K_{1}(a) = K_{1\;o} + \hat{K}_{1}(a)  ~,\ee
where
\be{n7} K_{1\;o} = \int_{2 \pi \alpha}^{2 \pi} J^{2}(y) y^{2} dy  ~\ee     
and 
\be{n8} \hat{K}_{1}(a) = \int_{\frac{2 \pi}{a}}^{2 \pi \alpha} J^{2}(y) y^{2} dy ~.\ee 
We find that $K_{1\;o}$ is an essentially $a$-independent constant for each value of $c$. $\alpha$ is chosen small enough that the modes in $\hat{K}_{1}(a)$ satisfy $y \ll 1$. In this case we can use the small $x$ form for the Bessel functions, 
\be{n9}  J_{\nu}\left(x\right) = \frac{1}{\Gamma(\nu + 1)} \left(\frac{x}{2} \right)^{\nu}\left(1  + O\left(\frac{x}{2}\right)^{2}\right)  ~.\ee 
Therefore 
\be{n10} J(y) \approx \tilde{c}_{1}  \frac{1}{\Gamma(1 - \frac{1}{2}\sqrt{4c + 9})} \left(\frac{y}{2}\right)^{- \frac{1}{2}\sqrt{4c + 9}}  + 
\tilde{c}_{2}  \frac{1}{\Gamma(1 + \frac{1}{2}\sqrt{4c + 9})} \left(\frac{y}{2}\right)^{\frac{1}{2}\sqrt{4c + 9}}    ~.\ee
For $y \ll 1$, 
\be{n11} J(y) \approx \tilde{c}_{1}  \frac{1}{\Gamma(1 - \frac{1}{2}\sqrt{4c + 9})} \left(\frac{y}{2}\right)^{- \frac{1}{2}\sqrt{4c + 9}} ~.\ee
Since $k/H > 2\pi$, we can approximate $\tilde{c}_{1}$ by 
\be{n12} \tilde{c}_{1} \approx - \frac{s_{-}}{\sin\left(\frac{\pi \sqrt{4c+9}}{2}\right)} \approx - \frac{ \sin\left(\frac{k}{H}\right)}{\sin\left(\frac{\pi \sqrt{4c+9}}{2}\right)} ~.\ee
Therefore 
\be{n13}  \hat{K}_{1}(a) = \frac{2^{\sqrt{4c+9}}}{\Gamma^{2}(1 - \frac{1}{2}\sqrt{4c + 9})\sin^{2}\left(
\frac{\pi \sqrt{4c+9}}{2}\right)   }
\int_{2\pi/a}^{2\pi\alpha} \sin^{2}\left(ay\right)y^{2 -\sqrt{4c + 9}} dy   ~.\ee 
Since $\sin^{2}(ay)$ is rapidly varying when $a \gg 1$, we can replace it with its average value $\sin^{2}(ay) = 1/2$. With this we find 
\be{n14} \hat{K}_{1}(a) = 
\frac{2^{\sqrt{4c+9} - 1}}{\Gamma^{2}(1 - \frac{1}{2}\sqrt{4c + 9}) \sin^{2}\left(
\frac{\pi \sqrt{4c+9}}{2}\right) }
\frac{\left(\frac{a}{2 \pi}\right)^{\sqrt{4c 
+ 9}-3}}{(\sqrt{4c + 9} - 3)} 
\left[1 - \left(\frac{1}{\alpha a}\right)^{\sqrt{4c + 9}-3} 
\right]  
  ~.\ee

Thus an approximation showing the $a$ dependence explicitly is 
\be{ax1} K_{1}(a) = A + B a^{\gamma}\left[1- \left( \frac{1}{\alpha a}\right)^{\gamma} \right]  ~,\ee
where $A \equiv K_{1\;o}$, $\gamma = \sqrt{4c + 9} -3$ and 
$B$ is the $a$-independent factor in \eq{n14}. We give the values of $A$, $B$, and $\gamma$ as a function of $c$ in Table 1. Using $\alpha = 0.02$, we find excellent agreement with the exact value of $K_{1}(a)$ for $a > 50$. (In practice we will be interested in $a \gg 50$.)

\begin{table}[h]
\begin{center}
\begin{tabular}{|c|c|c|c|}
 \hline $c$	&	$A$     &   $B$  &   $\gamma$ \\
\hline	$0.1$	&	$7.64$ &	$4.49$	&	$0.066$ \\   	
\hline	$0.2$	&	$7.77$ &	$2.12$	&	$0.130$ \\   	
\hline	$0.3$	&	$7.94$ & $1.34$	&	$0.194$	\\ 
\hline	$0.5$	& $8.26$ & $0.72$	&	$0.317$ \\   		
\hline	$0.75$	&	$8.78$ & $0.45$	&	$0.464$	\\   	
\hline	$1.0$	&	$9.51$ &  $0.29$	&	$0.606$	\\   		
\hline
 \end{tabular} 
 \caption{\footnotesize{Parameters for the approximation to $K_{1}(a)$ when $\alpha = 0.02$.} } 
 \end{center}
 \end{table}

\subsection{Power Spectrum} 

     At a given value of $a$, the $\phi_{1}$ and $\phi_{2}$ fields will evolve to mean values of the order of $\overline{\phi}$ in a given horizon volume. We can then make a field redefinition so that $\phi_{1}$ has non-zero mean value (the radial direction) while $\phi_{2}$ has only fluctuations about 
zero mean (the angular direction). Then  
\be{e36} \phi_{1} \approx \overline{\phi} + \delta \phi_{1}   ~\ee 
and
\be{e37} \phi_{2} \approx \delta \phi_{2}   ~.\ee 
The angular field fluctuation is then 
\be{e38}   \delta \theta = \frac{\delta \phi_{2}}{\overline{\phi} + \delta \phi_{1}} \approx \frac{\delta \phi_{2}}{\overline{\phi}} \left(1 - \frac{\delta \phi_{1}(\bfx)}{\overline{\phi}}\right)  ~. \ee
Thus $\delta \theta = \delta \phi_{2}/\overline{\phi}$ will give the mean isocurvature perturbation while the spatial modulation of the isocurvature perturbation is
due to $\delta \phi_{1} (\bfx)$, 
\be{e38a} \left|\frac{\Delta (\delta \theta)(\bfx)}{\delta \theta}\right|  \approx  \left| \frac{\delta \phi_{1}(\bfx)}{\overline{\phi}}\right|  ~.\ee

   The isocurvature perturbations are generated when $\delta \phi_{2}$ fluctuations exit the horizon. Thus the power spectrum is determined by the superhorizon fluctuations of
 $\delta \phi_{2}$. For superhorizon modes we can use the small $x$ expansion of the Bessel function as before. Therefore 
\be{n15} < \delta \phi_{2}^{2} > = \frac{1}{16 \pi H a^{3} \Gamma^{2}(1 - \frac{1}{2} \sqrt{4c+9})
\sin^{2}\left(
\frac{\pi \sqrt{4c+9}}{2}\right) 
} \int \left(\frac{k}{2 a H} \right)^{-\sqrt{4c + 9}} k^{2} dk \propto \int k^{3 -\sqrt{4c + 9}} \frac{dk}{k}    ~.\ee 
Therefore the spectral index of the isocurvature perturbation is 
\be{n16}  n_{iso} = 4 - \sqrt{4c + 9}   ~.\ee
This is less than 1 for all $c > 0$, corresponding to a red isocurvature power spectrum. We give the spectral index as a function of $c$ in Table 2.

\begin{table}[h]
\begin{center}
\begin{tabular}{|c|c|c|c|c|c|}
 \hline $c$	&  $0.1$	& $0.3$	& $0.5$	& $0.75$ & $1.0$	 \\	
\hline  $n_{iso}$ & $0.934$  & $0.806$  & $0.683$  & $0.536$ & $0.394$   \\   	
\hline
 \end{tabular} 
 \caption{\footnotesize{Spectral index of the isocurvature perturbation for $c$ in the range 0.1 to 1.0.} } 
 \end{center}
 \end{table}

\subsection{Hemispherical Asymmetry} 

      The hemispherical asymmetry of the isocurvature perturbation is due to the difference in $\delta \theta$ across the horizon at present. This is due to the spatial change in $\delta \phi_{1}/\overline{\phi}$ across the horizon when the observed Universe exits the horizon. 
To make this clear, consider the evolution of $\delta \phi_{1}/\overline{\phi}$ once the $\delta \phi_{1}$ fluctuation has wavelength larger than the horizon. Within a horizon volume, $\delta \phi_{1}/\overline{\phi}$ is approximately spatially constant and, since the evolution is linear during the tachyonic regime, this ratio remains constant as $\overline{\phi}$ grows. Therefore the modulation over the volume of the present Universe is the same for all subsequent $\delta \theta$ perturbations as they exit the horizon and become fixed in amplitude.  

    We therefore calculate the rms value of 
$\delta \phi_{1}(\bfx + \delta \bfx) -   \delta \phi_{1}(\bfx)$ for $|\delta \bfx|$ corresponding to the comoving horizon $(aH)^{-1}$ at a given value of $a$. 
The mean squared value is 
$$ <(\delta \phi_{1}(\bfx + \delta \bfx) -   \delta \phi_{1}(\bfx))^{2}> = <\delta \phi_{1}(\bfx + \delta \bfx)^{2}> 
+ <\delta \phi_{1}(\bfx)^{2}> - 2< \delta \phi_{1}(\bfx + \delta \bfx) \delta \phi_{1}(\bfx) > $$
\be{e16} = 2\left[ <\delta \phi_{1}(\bfx)^{2}> - < \delta \phi_{1}(\bfx + \delta \bfx) \delta \phi_{1}(\bfx) > \right] ~.\ee  
$<\delta \phi_{1}^{2}(\bfx)>$ is given by 
\be{n17} <\delta \phi_{1}^{2}(\bfx)> = \frac{1}{(2 \pi)^{3}} \int  |\phi_{\bfk}|^{2}  d^{3}k   ~.\ee 
The cross-correlation is given by 
\be{n18} < \delta \phi_{1}(\bfx + \delta \bfx) \delta \phi_{1}(\bfx) > = \frac{1}{(2 \pi)^{3}} \int |\phi_{\bfk}|^{2} e^{i \bfk.\delta \bfx} d^{3}k   ~.\ee 
Therefore
\be{n18a}  <(\delta \phi_{1}(\bfx + \delta \bfx) -   \delta \phi_{1}(\bfx))^{2}> = \frac{2}{(2 \pi)^{3}} \int |\phi_{\bfk}|^{2} \left(1 - e^{i \bfk.\delta \bfx} \right)  d^{3}k =  \frac{2}{(2 \pi)^{3}} \int |\phi_{\bfk}|^{2} \left(1 - \cos(\bfk.\delta \bfx) \right)  d^{3}k   ~.\ee 
$\delta \bfx$ is along the radius of a horizon volume centred at $\delta \bfx = 0$, with $|\delta \bfx| = (aH)^{-1}$. For different $\bfk$ directions with equal value of $k$, we can average over the directions, therefore we make the mild approximation that 
$\bfk.\delta \bfx = k|\delta \bfx|\cos(\sigma) \approx k|\delta \bfx|/2$, where $\sigma$ is the angle between $\bfk$ and $\delta \bfx$.   
Therefore, integrating over superhorizon modes, we obtain 
\be{n19}   <(\delta \phi_{1}(\bfx + \delta \bfx) -   \delta \phi_{1}(\bfx))^{2}> = \frac{2}{\pi^{2}} \int_{k_{min}}^{k_{max}}  |\phi_{\bfk}|^{2} \sin^{2}\left(\frac{k}{2aH}\right) k^{2} dk   ~.\ee  
Using \eq{n2a} and changing variable to $y = k/aH$, we find 
\be{n20}  \overline{ \Delta \phi}^{2} \equiv  <(\delta \phi_{1}(\bfx + \delta \bfx) -   \delta \phi_{1}(\bfx))^{2}> = \frac{H^{2}}{2 \pi}  K_{2}(a)   ~,\ee 
where 
\be{n21} K_{2}(a) = \int_{\frac{2 \pi}{a}}^{2 \pi} J^{2}(y) \sin^{2} \left( \frac{y}{2} \right) y^{2} dy  ~.\ee 
We find that $K_{2}(a)$ is constant with respect to $a$ for values of $a \gae 10$. Thus the spatial modulation of the field across a horizon radius becomes independent of time, whereas the mean field increases with time. In Table 4 we give the values of $K_2$ for $c$ in the range 0.1 to 1.  

\begin{table}[h]
\begin{center}
\begin{tabular}{|c|c|c|c|c|c|c|}
 \hline $c$	& 0.1     &   0.2  &  0.3 & 0.5 & 0.75 & 1.0 \\
\hline	$K_{2}$ &  $3.56$    & $3.58$	  & $3.59$  &	 $3.65$ & $3.72$  & $3.80$ \\   		
\hline
 \end{tabular} 
 \caption{\footnotesize{$K_{2}$ for values of $c$ in the range 0.1 to 1.0.} } 
 \end{center}
 \end{table}

\section{Results}    

    There are two cases of interest: (i) the possibility of observing isocurvature perturbations with a red spectrum and a significant hemispherical asymmetry\footnote{We calculate the hemispherical asymmetry in the rms magnitude of the isocurvature perturbation. The hemispherical isocurvature power asymmetry will then be a factor of 2 larger, since the power is proportional to the square of the perturbation.} and (ii) 
the specific case where a dark matter isocurvature perturbation with a large hemispherical asymmetry due to curvaton decay may account for the hemispherical power asymmetry observed by WMAP and Planck. 

\subsection{Isocurvature spectrum and hemispherical asymmetry} 

    In Figure 1 we show the rms spatial modulation of $\phi$ across the horizon, $\overline{\Delta \phi}/\overline{\phi}$, as a function of the number of e-foldings $\Delta N$ after the beginning of tachyonic evolution, for $c$ in the range 0.1 to 1. In Tables 4 and 5 we give the value of $\Delta N$ as a function of $c$ below which a modulation $\overline{\Delta \phi}/\overline{\phi}$ equal to 0.1 and 0.5 can be achieved.

   Rather generally, we find that a significant hemispherical asymmetry of the isocurvature perturbation can be achieved. Values of $\overline{\Delta \phi}/\overline{\phi} \gae 0.1$, corresponding to a greater than 10$\%$ hemispherical isocurvature perturbation asymmetry across the horizon radius, are obtained when 
$\Delta N < 100$ for $c = 0.1$ and $\Delta N < 15$ for $c = 1.0$. Therefore if our Universe exits the horizon when $\Delta N$ is within these limits, a greater than 10 $\%$ asymmetry in the isocurvature perturbation can be achieved. This asymmetry would be associated with a red isocurvature spectrum, therefore the combination of a hemispherical asymmetry and red spectrum would be a signature of this class of model. 

    In Figure 2 we show the value of $\overline{\phi}/H$ as a function of $\Delta N$ for $c$ in the range 0.1 to 1. In general, the value of $\overline{\phi}/H$ at which large $\overline{\Delta \phi}/\phi$ occurs is not larger than 7. Specifically, for $\overline{\Delta \phi}/\overline{\phi} = 0.1$ we find that $\overline{\phi}/H$ is approximately 7 for all $c$. Thus even though a large number of e-foldings of tachyonic growth can occur, the change in the mean field is relatively small. Therefore is quite plausible that the isocurvature field will not reach the minimum of its potential before the large asymmetry is generated.   

\subsection{WMAP/Planck hemispherical asymmetry} 

     In \cite{kam} it was proposed that the WMAP hemispherical asymmetry might be explained by a dark matter isocurvature perturbation due to a curvaton field with a large hemispherical asymmetry. In the case of an axion-like curvaton, this would require a modulation of the curvaton across the horizon radius by $\overline{\Delta \phi}/\overline{\phi} \gae 0.5$. 
From Figure 1 and Table 5, we see that it is possible to generate large modulations of the curvaton in this model. 
For $c = 0.1$, $\overline{\Delta \phi}/\overline{\phi} \geq 0.5$
is generated if 
$\Delta N < 40$, while for $c = 1$ this is achieved if $\Delta N < 10$. 

     A strong constraint is from quasar number counts, which show that the hemispherical asymmetry must decrease at large $l$ \cite{quasar}. This can be achieved if the asymmetry is due to an isocurvature perturbation, since the isocurvature power naturally decays relative to the adiabatic power at large $l$ \cite{kam}. In the curvaton model for dark matter isocurvature perturbations, the curvaton decays after dark matter freeze-out, producing equal adiabatic and isocurvature contributions to the total power spectrum, both of which have a hemispherical asymmetry. This leads to a strong constraint on the possible hemispherical asymmetry from the quasar bound, since the adiabatic asymmetry does not decrease with scale and must satisfy the quasar bound at large $l$. 
 
     However, the analysis of \cite{kam} assumes a scale-invariant curvaton spectrum. The effect of a red spectrum will be to decrease both the isocurvature and adiabatic power on larger scales. This should therefore relax the scale-dependence bound on the total hemispherical asymmetry, allowing a larger hemispherical asymmetry and so a smaller contribution to the total CMB power by the dark matter isocurvature perturbation.

\begin{figure}[htbp]
\begin{center}
\epsfig{file=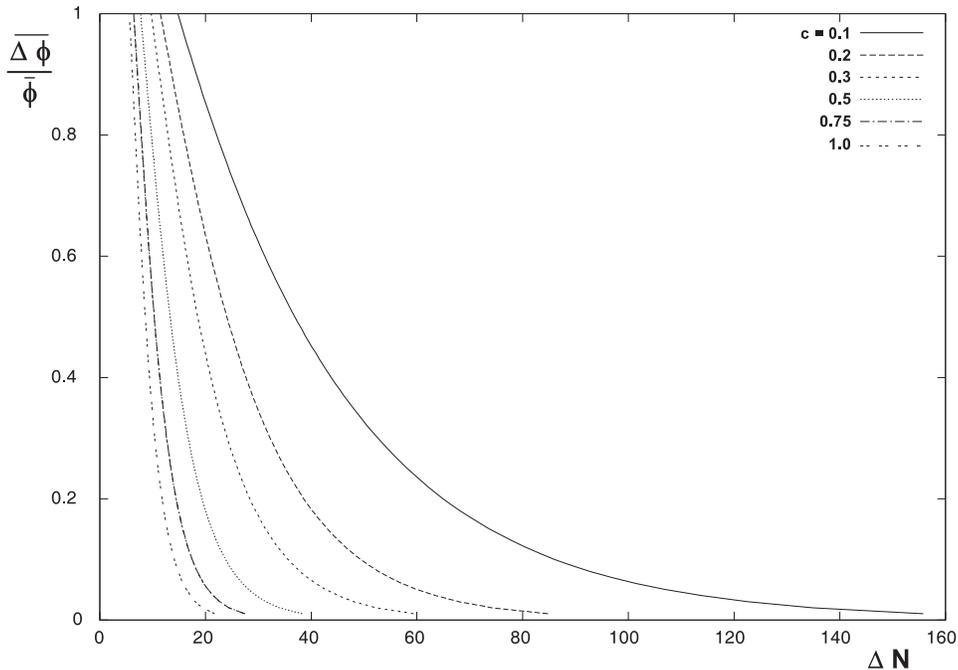, width=0.5\textwidth, angle = -90}
\caption{Values of $\overline{ \Delta \phi}/\overline{\phi}$ versus $\Delta N$ for $c$ is the range 0.1 to 1.0.}
\label{fig1}
\end{center}
\end{figure}

\begin{figure}[htbp]
\begin{center}
\epsfig{file=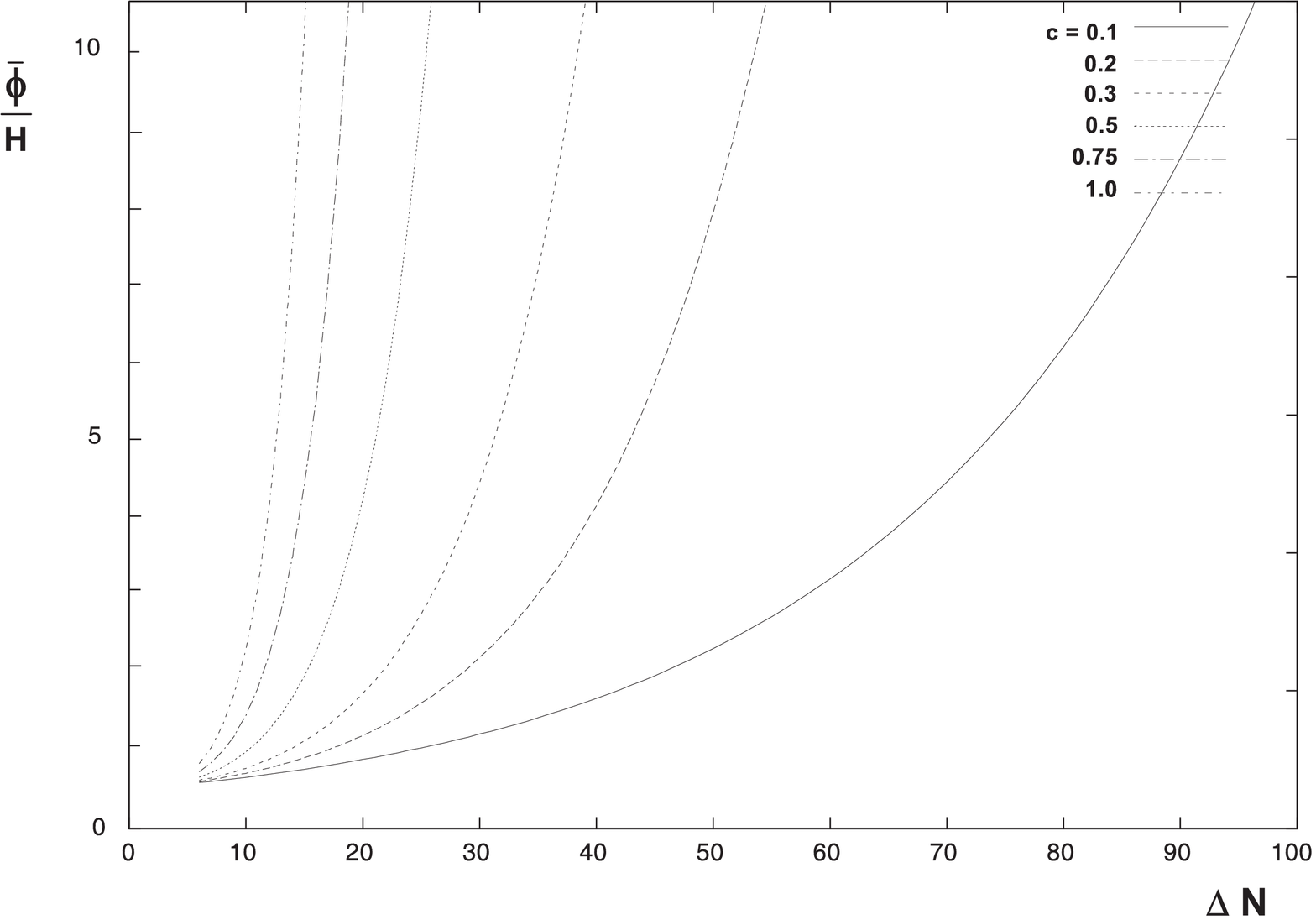, width=0.5\textwidth, angle = -90}
\caption{Values of $\overline{\phi}/H$ versus $\Delta N$ for $c$ in the range 0.1 to 1.0.}
\label{fig2}
\end{center}
\end{figure}

\begin{table}[h]
\begin{center}
\begin{tabular}{|c|c|c|}
 \hline $c$	 &  $\Delta N$ & $\overline{\phi}/H$ \\
\hline	$0.1$	&  $83.1$	&	$6.84$ \\   	
\hline	$0.3$	&  $34.6$	&	$6.86$ \\   	
\hline	$0.5$	&  $23.2$	& $6.94$	\\ 
\hline	$0.75$ & $17.1$      & $7.06$ \\ 
 \hline	$1.0$	 & $13.8$ & $7.10$	\\   	
\hline
 \end{tabular} 
 \caption{\footnotesize{Number of e-foldings of tachyonic growth and the mean field when $\overline{\Delta \phi}/\overline{\phi} = 0.1$.} } 
 \end{center}
 \end{table}

\begin{table}[h]
\begin{center}
\begin{tabular}{|c|c|c|}
 \hline $c$	&  $\Delta N$ & $\overline{\phi}/H$ \\
\hline	$0.1$	& $36.8$	&	$1.50$ \\   	
\hline	$0.3$	& $18.6$	&	$1.52$ \\   	
\hline	$0.5$	& $13.3$	& $1.52$	\\ 
\hline	$0.75$ &	$10.3$      & $1.54$ \\  
\hline	$1.0$	 & $8.6$ & $1.54$	\\   	
\hline
 \end{tabular} 
 \caption{\footnotesize{Number of e-foldings of tachyonic growth and the mean field when $\overline{\Delta \phi}/\overline{\phi} = 0.5$.} } 
 \end{center}
 \end{table}

\section{Conclusions} 

           We have studied the properties of the isocurvature perturbation generated by a complex field which grows from $\Phi = 0$ in a tachyonic potential. The model has one free parameter, the tachyonic mass term, $cH^2$.   

   A general feature of this class of model is that the spectral index of the isocurvature perturbations is less than one, with the spectrum becoming increasingly red as $c$ increases, with $n_{iso} = 0.934$ for $c  = 0.1$ and $n_{iso} = 0.394$ for $c = 1.0$. 

   We find that a hemispherical isocurvature perturbation asymmetry is a natural feature of isocurvature perturbations in this model. Asymmetries of the isocurvature perturbation greater than 10$\%$ 
across the horizon can be easily achieved if the observed Universe exits the horizon during inflation at between 14 and 80 e-foldings after the beginning of tachyonic evolution for $c$ in the range 0.1 to 1.0. 

  Therefore a signature of this class of isocurvature perturbation model would be the observation of a red isocurvature spectrum together with a hemispherical isocurvature perturbation asymmetry. 

      A topical application of this model is to the hemispherical power asymmetry of the CMB, first observed by WMAP and recently confirmed by Planck. It has been proposed that this might be explained by a dark matter isocurvature perturbation due to the decay of a curvaton which has a spatial modulation across the present horizon radius of at least 50$\%$ \cite{kam}. 

Our analysis of isocurvature perturbations can be directly applied to the case of an axion-like curvaton field. We find that large hemispherical asymmetries of the dark matter isocurvature perturbation can easily be achieved. For the case $c = 0.1$, a 50$\%$ modulation of the curvaton field requires that the observed Universe exits the horizon no later than 37 e-foldings after the beginning of tachyonic evolution, while for $c = 1.0$ it must exit before 9 e-foldings.

    A possible advantage of the tachyonic model is that the red spectrum can additionally suppress the isocurvature power and hemispherical asymmetry at large $l$. In the analysis of \cite{kam}, an important constraint comes from quasar number counts and other observations, which show that the hemispherical power asymmetry is smaller at larger $l$. In the case where the dark matter isocurvature perturbation comes from curvaton decay, there is an equal adiabatic perturbation which also has a hemispherical asymmetry. Since, unlike the isocurvature perturbation,  this does not decay at large $l$, the quasar bound limits the amount of adiabatic hemispherical asymmetry (and so dark matter isocurvature asymmetry) for a given adiabatic perturbation from curvaton decay \cite{kam}. The additional suppression due to the red spectrum should therefore relax the quasar bound and allow a larger hemispherical asymmetry. This could in turn relax constraints from the upper bound on the isocurvature perturbation.

   It remains to be established whether an isocurvature perturbation from curvaton decay can consistently account for the WMAP/Planck hemispherical asymmetry while satisfying the latest Planck constraints on the isocurvature perturbation \cite{pesky}. Our model shows that a large hemispherical asymmetry in the isocurvature perturbation can be generated in a plausible model. It also illustrates the need for any future analysis to include a possible red spectrum for the isocurvature and curvaton perturbations.

 \section*{Acknowledgements}

The work of JM is supported by the Lancaster-Manchester-Sheffield Consortium for Fundamental Physics under STFC grant
ST/J000418/1.

\renewcommand{\theequation}{A-\arabic{equation}}
 \setcounter{equation}{0} 

\section*{Appendix. Wigner function method for semi-classical tachyonic mode evolution}

  We use the same method used in numerical studies of tachyonic preheating and oscillon formation \cite{felder,mbs,latticeeasy}. The original formulation is given in \cite{staro}. A summary of the methods is contained in \cite{latticeeasy} and in the Appendix of \cite{mbs}.    

    For consistency with \cite{latticeeasy,mbs}, we will consider a finite box of volume $V$ and side $L$, taking the continuum limit where appropriate. The modes are defined using periodic boundary conditions, with wavenumbers $k_{i} = 2 \pi n_{i}/L$. The field $\Phi$ can be expanded in terms of real fields,   $\Phi = (\phi_{1} + i \phi_{2})/\sqrt{2}$. Since the evolution 
is linear during the tachyonic regime, it is the same for $\phi_{1}$ and $\phi_{2}$, so we will consider a generic real scalar field $\phi$. The action is
\be{e2} S = \int d^{4}x \sqrt{-g} \left[ \frac{1}{2} \partial_{\mu} \phi \partial^{\mu} \phi + \frac{c H^{2}}{2} \phi^{2} \right]   ~.\ee
This can be expressed in terms of $y = a\phi$ and integrated over comoving coordinates and conformal time 
\be{e3}   S  =  \int d^{4}x \left[\frac{1}{2}\left(y^{'} - \frac{a^{'}}{a}y \right) - \frac{\left(\nabla y\right)^{2}}{2} +\frac{c H^2 a^2}{2} y^{2} \right]        ~,\ee
where primes denote derivatives with respect to $\eta$.

This is canonically quantized in terms of $y$ and its conjugate momentum $\pi = \partial{\cal L}/\partial y$. The mode expansion of the fields is 
\be{e4} y(\bfk, \eta) = \frac{1}{\sqrt{V}} \sum_{\bfk} q(\bfk, \eta) e^{i \bfk.\bfx}    ~\ee
and
\be{e5} \pi(\bfk, \eta) = \frac{1}{\sqrt{V}} \sum_{\bfk} p(\bfk, \eta) e^{i \bfk.\bfx}    ~,\ee
where 
\be{e6}  p(\bfk, \eta) = q^{'}(\bfk, \eta)  - \frac{a^{'}}{a} 
q(\bfk, \eta)   ~.\ee 
These are quantized via 
\be{e7} q(\bfk, \eta) = f_{\bfk} a_{\bfk} + f^{\dagger}_{\bfk} a_{-\bfk}^{\dagger}      ~\ee
and
\be{e8} p(\bfk, \eta) = -i g_{\bfk} a_{\bfk} + i g^{\dagger}_{\bfk} a_{-\bfk}^{\dagger}      ~,\ee
where $[a_{\bfk}, a^{\dagger}_{\bfl}] = \delta_{\bfk,\bfl}$. The mode functions satisfy the classical field equation 
\be{e9}   f_{\bfk}^{''} - \frac{a^{''}}{a} f_{\bfk} = -( \bfk^{2} + m^{2} a^{2} )  f_{\bfk} ~.\ee
In the case of de Sitter space with constant $H$, and defining $\eta = -1/aH$, this becomes 
\be{e10}   f_{\bfk}^{''} + \left[k^{2} - \frac{\left(2 - \frac{m^{2}}{H^{2}}\right)}{\eta^{2}} \right] f_{\bfk}  = 0 ~.\ee 
The correctly normalized solution for quantum modes in de Sitter space is then
\be{e11} f_{\bfk}(\eta) = \frac{\sqrt{\pi
}}{2} e^{i\left( \frac{\pi}{2} \nu + \frac{\pi}{4} \right)} \sqrt{-\eta} H_{\nu}^{(1)}(-k \eta)   ~,\ee
where $\nu^2 = 9/4 - m^{2}/H^{2} $.  In the following we will consider the field to be effectively massless on subhorizon scales prior to the tachyonic growth era, therefore 
\be{e12} f_{\bfk} = i \frac{\sqrt{\pi}}{2} \sqrt{-\eta} H_{\frac{3}{2}}^{(1)} (-k \eta) = \frac{e^{-ik\eta}}{\sqrt{2k}} \left[1 - \frac{i}{k \eta}\right]   ~\ee 
and
\be{e13} g_{\bfk} \equiv  i \left( f_{\bfk}^{'} - \frac{a^{'}}{a} f_{\bfk} \right)  = \sqrt{\frac{k}{2}} e^{-ik\eta}    ~.\ee

 The Wigner function method replaces the quantum modes by classical modes at an initial time. The Wigner function is a quasi-probability distribution for quantum states, $W(q,p,t)$, which gives the probability for $q$ on integrating over $p$ and vice-versa. In the semi-classical limit the Wigner function $W(q,p,t)$ becomes a classical phase space distribution. However, we can define the classical phase distribution for the late-time semi-classical state at any time during the linear evolution of the fields, even when the semi-classical limit is not valid. This is because taking $\hbar \rightarrow 0$ does not affect the late-time semi-classical state. Therefore if we take $\hbar \rightarrow 0$ at earlier times, the resulting classical phase space distribution must reproduce the correct semi-classical limit at late times. The only requirement is that the quantum fields must enter the semi-classical regime during the linear evolution of the fields, since it is only during linear evolution that we can evaluate the classical probability distribution from the Wigner function. We also note that this is an exact solution of the field theory in the semi-classical limit; there is no use of perturbation theory. 

  The resulting classical modes are Gaussian distributed with a random phase. The distribution is given by \cite{latticeeasy,mbs}   
\be{a1} P(q_{\bfk},t) = \frac{1}{\pi^{2}}\exp\left[-\frac{\left|q_{\bfk}^{2}\right|}{\left|f_{\bfk}\right|^{2}}\right]     ~,\ee 
with $p_{\bfk} = F_{\bfk} q_{\bfk}/|f_{\bfk}|^{2}$ and 
$F_{\bfk} = Im(f_{\bfk}^{\dagger}g_{\bfk}) = 1/2k\eta$. Thus the rms 
classical modes at conformal time $\eta$ are 
\be{a2} |q(\bfk,\eta)|_{rms} = \frac{1}{\sqrt{2k}} \left(1 + 
\frac{1}{k^{2} \eta^{2}} \right)^{1/2}    ~\ee
and
\be{a2} |p(\bfk,\eta)|_{rms} = \frac{1}{\sqrt{2k}|\eta|} \left(1 + 
\frac{1}{k^{2} \eta^{2}} \right)^{-1/2}    ~.\ee
These give the initial conditions at $t = 0$, 
 corresponding to $\eta = -1/H$, for the classical evolution of the modes in the tachyonic era.


\end{document}